\documentclass[pra,aps,twocolumn,showpacs,floatfix]{revtex4}

\usepackage{amsmath}
\usepackage{bm}
\usepackage{graphicx}

\begin{document}

\title{Bose-Einstein droplet in free space}

\author{Hiroki Saito}
\author{Masahito Ueda}
\affiliation{Department of Physics, Tokyo Institute of Technology,
Tokyo 152-8551, Japan \\
and CREST, Japan Science and Technology Corporation (JST), Saitama
332-0012, Japan
}

\date{\today}

\begin{abstract}
We show that a droplet of a Bose-Einstein condensate can be dynamically
stabilized in free space by rapid oscillations of interatomic interactions
between attractive and repulsive through, e.g., the Feshbach resonance.
Energy dissipation, which is present in realistic situations, is found to
play a crucial role to suppress dynamical instabilities inherent in
nonlinear nonequilibrium systems.
\end{abstract}


\maketitle

\section{Introduction}

Matter-wave bright solitons --- stable solitary waves whose density
is greater than the background one --- have recently been realized in a
quasi one-dimensional (1D) Bose-Einstein condensate
(BEC)~\cite{Khay,Strecker}.
In 1D, the quantum kinetic pressure counterbalances an attractive
interatomic interaction, allowing a stable bright soliton to be formed.
However, in 2D or higher dimensions, bright solitons are always unstable
against collapse or expansion as long as the attractive interaction is
constant in time~\cite{Saito}.

It has been demonstrated by Abdullaev {\it et al.}~\cite{Ab03} and by
us~\cite{Saito} that by rapid oscillations of interaction between
attractive and repulsive through, e.g., the Feshbach
resonance~\cite{Inouye,Cornish}, a BEC can be stabilized in a quasi 2D
space without a radial trapping potential.
The mechanism of the stabilization is similar to that of an inverted
pendulum, in which a bob is stabilized above the vertically oscillating
pivot~\cite{Landau,Feynman}.
However, the stabilization of a BEC in 3D free space has been elusive due
to complex dynamical instabilities inherent in nonlinear nonequilibrium
systems.

In the present paper, we show that a droplet~\cite{note} of a BEC can be
stabilized in 3D free space by using oscillating interactions.
In order to simulate experimental situations at finite temperature, we
take into account the effect of dissipation caused by the interaction
between a BEC and a thermal cloud.
This can be incorporated in mean field theory by a non-unitary term
added to the Gross-Pitaevskii (GP) equation.
This model has successfully been used to describe damping of collective
modes~\cite{Choi} and vortex nucleation~\cite{Tsubota}.
We find that the dissipation plays a crucial role in the stabilization of
a BEC droplet by suppressing the dynamical instabilities, and show that
the droplet can indeed be stabilized for a realistic amount of
dissipation.

This paper is organized as follows.
Section~\ref{s:GP} introduces the GP equation with oscillating
interactions, and shows that the dependence of the dynamics on the
oscillation frequency can be eliminated by the scaling property when a
trapping potential vanishes.
Section~\ref{s:num} presents the main results of this paper, in which a
BEC droplet is shown to be stabilized in 3D free space by use of
oscillating interactions.
The stability diagrams with respect to the interaction and dissipative
parameters are also obtained.
Section~\ref{s:var} studies the variational analysis, and
Sec.~\ref{s:conc} concludes this paper.

\section{The scaling property of the Gross-Pitaevskii equation}
\label{s:GP}

We consider the GP equation in the presence of dissipation
given by~\cite{Choi,Tsubota}
\begin{eqnarray} \label{GP}
(i - \gamma) \hbar \frac{\partial \psi}{\partial t} & = &
-\frac{\hbar^2}{2 m} \nabla^2 \psi + V_{\rm trap}({\bf r}, t) \psi +
\frac{4 \pi \hbar^2 a(t)}{m} |\psi|^2 \psi \nonumber \\
& & + i \gamma \mu \psi,
\end{eqnarray}
where $\gamma$ is a phenomenological dissipation constant which is to be
determined experimentally, and the trapping potential $V_{\rm trap}({\bf
r}, t)$ and the s-wave scattering length $a(t)$ are controlled to vary in
time.
The last term on the right-hand side of Eq.~(\ref{GP}) guarantees the
normalization $\int d{\bf r} |\psi|^2 = N$ with $N$ being the number of
atoms, where the chemical potential is given by
\begin{equation}
\mu = \int d{\bf r} \psi^* \left( -\frac{\hbar^2}{2 m} \nabla^2 + V_{\rm
trap}({\bf r}, t) + \frac{4 \pi \hbar^2 a(t)}{m} |\psi|^2 \right) \psi.
\end{equation}

For simplicity, we assume an isotropic trapping potential
\begin{equation}
V_{\rm trap}({\bf r}, t) = f_{\rm trap}(t) \frac{m \omega^2}{2} r^2,
\end{equation}
where $f_{\rm trap}(t)$ is unity at $t = 0$, then decreases towards zero,
and vanishes at some time.
The s-wave scattering length is made to oscillate as
\begin{equation}
a(t) = f_{\rm int}(t) (a_0 + a_1 \sin \Omega t),
\end{equation}
where $f_{\rm int}(t)$ ramps up from 0 to 1.
The ramp functions $f_{\rm trap}(t)$ and $f_{\rm int}(t)$ are introduced
to avoid initial nonadiabatic disturbances that cause dynamic
instabilities.
These functions must be chosen so that neither collapse nor expansion
occurs in the course of the ramping.

Normalizing the length, time, energy, and wave function by $(\hbar / m
\Omega)^{1/2} \equiv d_0$, $\Omega^{-1}$, $\hbar \Omega$, and $\sqrt{N}
d_0^{-3/2}$, respectively, we obtain the normalized form of the GP
equation:
\begin{equation} \label{GPn}
(i - \gamma) \frac{\partial \psi}{\partial t} = -\frac{\nabla^2}{2} \psi +
\frac{f_{\rm trap}(t)}{2} \left( \frac{\omega}{\Omega} \right)^2 r^2 \psi
+ g(t) |\psi|^2 \psi + i \gamma \mu \psi,
\end{equation}
where $g(t) \equiv 4\pi N a(t) / d_0 \equiv f_{\rm int}(t) (g_0 + g_1 \sin
t)$, and the wave function is normalized as $\int d{\bf r} |\psi|^2 = 1$.
It should be noted that once the trapping potential is switched off
[$f_{\rm trap}(t) = 0$], Eq.~(\ref{GPn}) no longer depends on $\Omega$
explicitly.
This implies that the $\Omega$ dependence can be eliminated by
appropriate scaling of the time, length, and interaction.
The dynamics at driving frequency $\alpha \Omega$ can be reduced to those
at $\Omega$ by rescaling $t \rightarrow \alpha t$, ${\bf r}
\rightarrow \sqrt{\alpha} {\bf r}$, and $a(t) \rightarrow \sqrt{\alpha}
a(t)$.
For example, if large $\Omega$ is unavailable experimentally, we have only 
to increase the strength of interaction and enlarge the size of the BEC
droplet.

\section{Numerical simulations}
\label{s:num}

We numerically solve the GP equation (\ref{GPn}) using the Crank-Nicholson
scheme~\cite{Ruprecht}, where the following simple forms of the ramp
functions are used:
\begin{eqnarray}
\label{ftrap}
f_{\rm trap}(t) & = & \left\{ \begin{array}{ll} 1 - t / T_{\rm trap} & (0
\leq t \leq T_{\rm trap}) \\ 0 & (t > T_{\rm trap}), \end{array}
\right. \\
\label{fint}
f_{\rm int}(t) & = & \left\{ \begin{array}{ll} 1 - (t / T_{\rm int} -
1)^2 & (0 \leq t \leq T_{\rm int}) \\ 1 & (t > T_{\rm int}). \end{array}
\right.
\end{eqnarray}
We use a quadratic function for $f_{\rm int}(t)$ because it makes initial
nonadiabatic disturbances smaller than linear ramps.

\begin{figure}[tb]
\includegraphics[width=8.4cm]{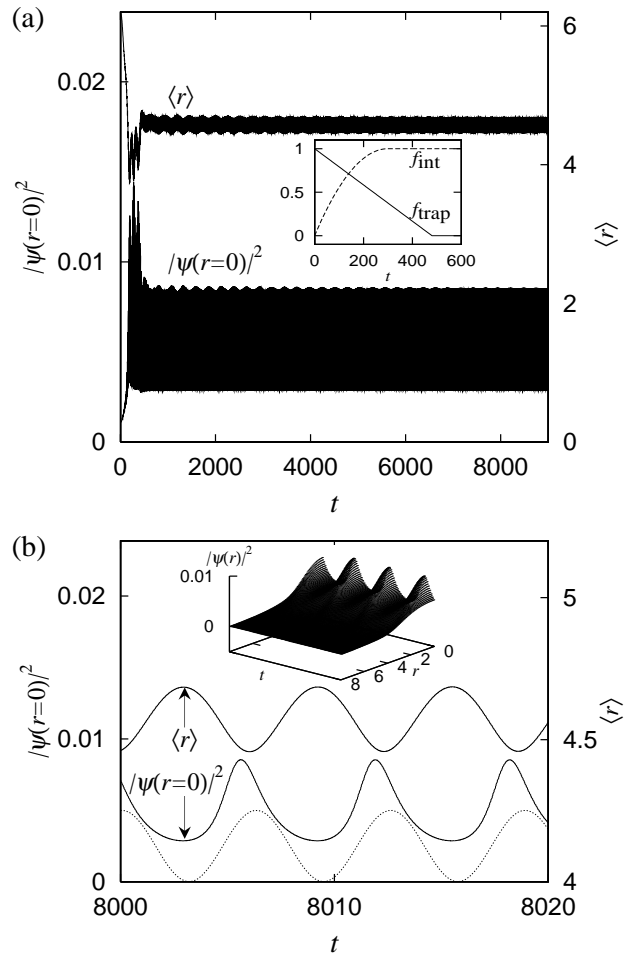}
\caption{
(a) Time evolution of the peak density $|\psi(r=0)|^2$ (left axis) and
monopole moment $\langle r \rangle = \int d{\bf r} r |\psi|^2$ (right
axis) for the oscillating interaction $g(t) = -69 + 155 \sin t$ with
$\gamma = 0.3$.
The initial state is the noninteracting ground state in a trapping
potential $r^2 / 1800$.
Then the interaction is gradually turned on, and the trapping potential
is simultaneously turned off as shown in the inset.
(b) A magnification of (a).
The inset shows the density profile $|\psi(r)|^2$ from $t = 8000$ to $t =
8020$.
The dotted line is a sine function for comparison of the phase.
}
\label{f:stable}
\end{figure}

Figure~\ref{f:stable} shows the time evolution of the peak density
$|\psi(r = 0)|^2$ and the monopole moment $\langle r \rangle = \int d{\bf
r} r |\psi|^2$ for $g_0 = -69$ and $g_1 = 155$.
The dissipation constant $\gamma$ is taken to be 0.03~\cite{Choi,Tsubota},
and the ramp parameters are $T_{\rm trap} = 16$ and $T_{\rm int} = 10$.
The initial state is chosen to be the noninteracting ground state for a
trapping potential with $\omega = \Omega / 30$.
The density and width oscillate at the driving frequency. [The black bands
in Fig.~\ref{f:stable} (a) represent rapid oscillations of $\langle r
\rangle$ and $|\psi(0)|^2$ which are beyond the resolution limit.]
We note that the amplitudes and mean values of both $\langle r \rangle$
and $|\psi(0)|^2$ converge to constant values, indicating that a BEC
droplet is dynamically stabilized.
The initial disturbances last for a relatively long time shown as slow
oscillations or ripples of $\langle r \rangle$ and $|\psi(0)|^2$ in
Fig.~\ref{f:stable} (a), which gradually decay due to dissipation.
From the time evolution of the density profile $|\psi(r)|^2$ [inset in
Fig.~\ref{f:stable} (b)], we find that the droplet exhibits breathing-mode
oscillations with a large amplitude.
This indicates that large oscillations of mass flow persist even after
stabilization.
We find from Fig.~\ref{f:stable} (b) that the phase of the breathing-mode
oscillations of the droplet lags slightly behind that of the oscillating
interaction shown as the dotted curve.
This retardation is considered to be due to dissipation because there is
no phase difference in the dissipation-free 2D case~\cite{Saito}, where
$|\psi(0)|^2$ and $\langle r \rangle$ are in-phase and out-of-phase with
the oscillating interaction.

\begin{figure}[tb]
\includegraphics[width=8.4cm]{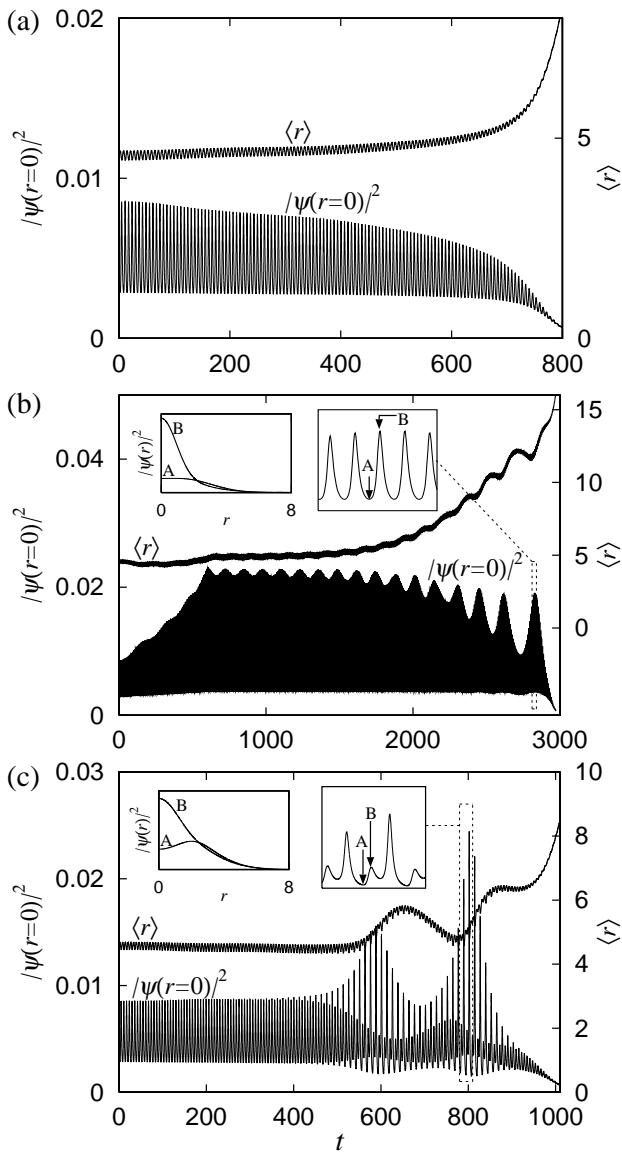}
\caption{
Time evolution of the peak density $|\psi(r=0)|^2$ (left axis) and
monopole moment $\langle r \rangle = \int d{\bf r} r |\psi|^2$ (right
axis).
The initial state is prepared as in Fig.~\protect\ref{f:stable} (a), and
$(g_0, g_1)$ are linearly changed from $(-69, 155)$ to $(-69, 160)$ in (a),
$(-69, 129)$ in (b), and $(-72, 173)$ in (c) during $0 \leq t \leq 600$.
The insets in (b) and (c) are magnifications of $|\psi(r=0)|^2$ in the
dashed squares and the density profiles $|\psi(r)|^2$ at the times
indicated by A and B.
}
\label{f:unstable}
\end{figure}

We gradually changed $g_0$ and $g_1$ from those of the stable droplet
state in Fig.~\ref{f:stable}, and found distinct types of instabilities
as shown in Fig.~\ref{f:unstable}.
When $|g_0|$ is decreased, a droplet expands because of a decrease in an
attractive interaction as shown in Fig.~\ref{f:unstable} (a).
The expansion occurs also by an increase in $g_1$, since the effective
repulsive interaction due to the oscillation is proportional to $g_1^2$
[see Eq.~(\ref{eom2})].
When $|g_0|$ is increased and $g_1$ is decreased, the system becomes
dynamically unstable against the slow oscillation and eventually expands
away as shown in Fig.~\ref{f:unstable} (b).
The expansion occurs, despite the fact that an increase in $|g_0|$
enhances the attraction and a decrease in $g_1$ suppresses the repulsion.
The slow oscillation is seen in the profile of the maxima of $|\psi(0)|^2$
(the ripple in the upper edge), while the minima is almost constant.
Figure~~\ref{f:unstable} (c) shows an instability that arises for large
$|g_0|$ and $g_1$.
Since the modulation at half frequency grows, this instability is
considered to arise from parametric resonance.
The density profiles in the inset show that higher radial modes are
excited, which leads to destabilization of the droplet.

\begin{figure}[tb]
\includegraphics[width=8.4cm]{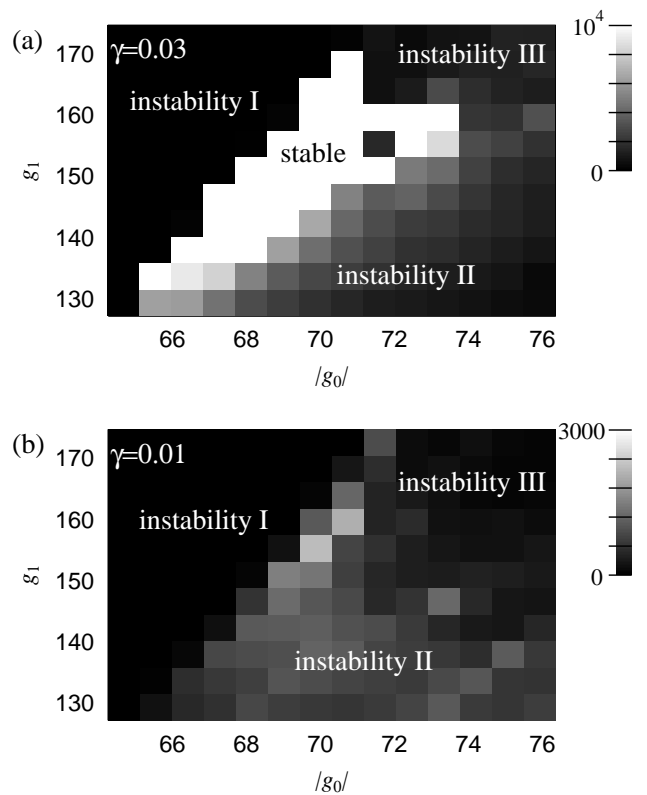}
\caption{
The lifetime of a BEC droplet after a stable state is prepared as in
Fig.~\protect\ref{f:stable}, and the interaction is linearly changed during
$0 \leq t \leq 600$ as in Fig.~\protect\ref{f:unstable}.
(a) The black region corresponds to the one in which a droplet expands
during the change of the interaction, and the white region corresponds to
the one in which a droplet survives until at least $t = 10600$.
The behaviors in the regions ``instability I, II, and III'' correspond to
those in Fig.~\protect\ref{f:unstable} (a), (b), and (c), respectively.
(b) The parameter $\gamma$ is also linearly changed as well as the
interaction during $0 \leq t \leq 600$ from $\gamma = 0.03$ to $0.01$.
}
\label{f:contour}
\end{figure}

We prepare a stable droplet in the same manner as in Fig.~\ref{f:stable},
and change the parameters to various values to obtain the stability
diagram shown in Fig.~\ref{f:contour}.
The parameters $g_0$ and $g_1$ are changed linearly during $0 \leq t \leq
600$.
We define the lifetime of the droplet as the duration between $t = 600$
and the time at which the droplet begins to expand indefinitely.
The regions referred to as ``instability I, II, and III'' in
Fig.~\ref{f:contour} correspond to the instabilities demonstrated in
Fig.~\ref{f:unstable} (a), (b), and (c).
We find from Fig.~\ref{f:contour} (a) that the stable region is surrounded
by the three distinct types of instabilities.
The boundary between the ``stable'' and ``instability I'' regions are
sharp while the ``instability II'' gradually sets in as $|g_0|$ increases
and $g_1$ decreases.
The collapse occurs for large $|g_0|$ and small $g_1$, but the
corresponding parameter regime lies outside Fig.~\ref{f:contour}.

In Fig~\ref{f:contour} (b), $\gamma$ is linearly decreased from 0.03 to
0.01 as $|g_0|$ and $g_1$ are (also linearly) changed.
The instability regions enlarge as compared with those of
Fig~\ref{f:contour} (a), and the stability region disappears.
From this result, we conclude that the stability region exists only for
$\gamma \neq 0$.
However, we cannot exclude the possibility that other stable states exist,
since we do not investigate the entire functional space.

In numerical calculations of the time evolution, we must pay special
attention to the boundary effect.
Since the trapping potential is absent, the atoms that escape from the
droplet spread out, reflect at the boundary, and return to the droplet
region, producing spurious boundary effects.
For example, when the Dirichlet boundary condition is imposed at $r =
116$, the lifetime of the droplet (whose definition is the same as
in Fig.~\ref{f:contour}) for $g_0 = -69$ and $g_1 = 134$ becomes $\simeq
4100$, while the correct value is $\simeq 3600$.
Therefore, the spatial cutoff in the numerical analysis must be much
larger than the size of the droplet.

\begin{figure}[tb]
\includegraphics[width=8.4cm]{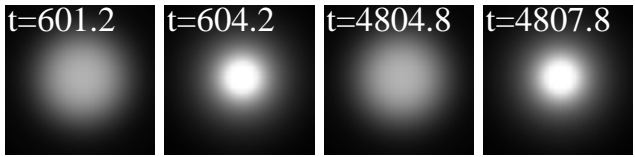}
\caption{
The profiles of the column density of a BEC droplet obtained by full 3D
numerical calculations.
Since the isotropic symmetry is preserved, we obtain the same column
density from any direction.
The size of the images is $12 \times 12$ in units of $(\hbar / m
\Omega)^{1/2}$.
}
\label{f:3D}
\end{figure}

The numerical simulations that we have presented so far have been carried
out under the assumption of the spherical symmetry of the system.
However, this geometry cannot generate multipole dynamical instabilities
which destroy a BEC droplet.
In order to check the absence of such multipole instabilities, we have
performed full 3D calculations by discretizing the space of $100 \times
100 \times 100$ size in our dimensionless unit into a $256 \times 256
\times 256$ mesh.
A limited spatial size due to our computational power causes spurious
boundary effects and a rough mesh produces numerical errors.
Nevertheless, as shown in Fig.~\ref{f:3D}, snapshots of time evolution of
a BEC droplet appear isotropic and do not show any multipole instability.
We have thus confirmed that a BEC droplet can be stabilized without
multipole instabilities.

\section{Variational analysis}
\label{s:var}

The Gaussian variational wave function well describes the dynamical
stabilization of BECs in 2D free space~\cite{Saito,Ab03}.
We examine a variational method in 3D using the Gaussian trial
function as~\cite{Garcia,SaitoA}
\begin{equation} \label{gauss}
\psi_{\rm var}({\bf r}, t) = \frac{1}{\pi^{3/4} R^{3/2}} \exp\left(
-\frac{r^2}{2 R^2} + i \frac{\dot{R} r^2}{2 R} \right),
\end{equation}
where $R(t)$ is the variational parameter that characterizes the size of
the condensate.
Substituting Eq.~(\ref{gauss}) into the action
\begin{equation} \label{S}
S = \int dt d{\bf r} \psi^* \left( -i \frac{\partial}{\partial t} -
\frac{\nabla^2}{2} + \frac{g}{2} |\psi|^2 \right) \psi,
\end{equation}
we obtain the equation of motion for $R$ as
\begin{equation} \label{eom}
\ddot{R} = -\frac{d}{d R} \left( \frac{1}{2 R^2} +
\frac{G}{6 R^3} \right),
\end{equation}
where $G(t) \equiv (g_0 + g_1 \sin t) / (2^{1/2} \pi^{3/2}) \equiv G_0 +
G_1 \sin t$.

We separate $R$ into the slowly varying part $R_0$ and the rapidly
oscillating part $\rho$ as $R = R_0 + \rho$.
According to Ref.~\cite{Landau}, an effective potential for $R_0$
is given by $f_1^2 / (4 \Omega^2)$, where $f_1$ is the amplitude of the
oscillating ``force'' for $R_0$ and $\Omega$ is its frequency.
From Eq.~(\ref{eom}), $f_1$ corresponds to $G_1 / (2 R_0^4)$.
Since we are using a system of units in which $\Omega = 1$, the effective
potential becomes $G_1^2 / (16 R_0^8)$, and then the equation of motion
for $R_0$ reads
\begin{equation} \label{eom2}
\ddot{R}_0 = -\frac{d}{d R_0} \left( \frac{1}{2 R_0^2} + \frac{G_0}{6
R_0^3} + \frac{G_1^2}{16 R_0^8} \right) \equiv -\frac{d U_{\rm eff}}{d
R_0}.
\end{equation}
This equation agrees with that in Ref.~\cite{Ab03} when $\Omega$ is much
larger than the characteristic frequencies of the system.
Thus, the oscillating interaction blocks the collapse by the effective
potential proportional to $R_0^{-8}$.

The effective potential $U_{\rm eff}$ has a local minimum for appropriate
values of $G_0$ and $G_1$.
For instance, when $G_0 = -70 / (2^{1/2} \pi^{3/2})$ and $G_1 = 140 /
(2^{1/2} \pi^{3/2})$, $U_{\rm eff}$ has a local minimum at $R_0 \simeq
2.4$.
In fact, by numerically solving the equation of motion (\ref{eom}) for
$R$, we find that a stable solution exists.
However, Fig.~\ref{f:contour} (b) implies that the system is unstable
against the dynamical instability for $\gamma = 0$.
This indicates that the simple Gaussian function (\ref{gauss}) is
insufficient to describe the instability of the system.

\begin{figure}[tb]
\includegraphics[width=8.4cm]{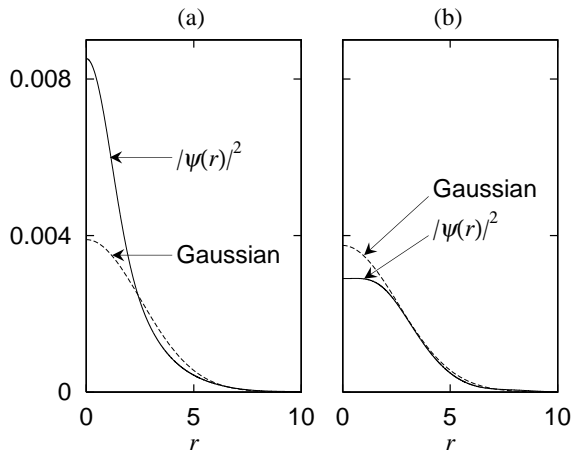}
\caption{
The density profiles $|\psi(r)|^2$ (solid curves) of the stable droplet
for $g_0 = -69$ and $g_1 = 155$ ($t \protect\gtrsim 8000$ in
Fig.~\protect\ref{f:stable}) when the peak density becomes (a) maximal and
(b) minimal.
The dashed curves are the Gaussian functions fitted to the density
profiles.
}
\label{f:diff}
\end{figure}

Figure~\ref{f:diff} illustrates the difference between the density
profiles of the stable droplet ($t \gtrsim 8000$ in Fig.~\ref{f:stable})
and the Gaussian functions fitted to them.
We adopt the least squares fitting of $r^2 |\psi(r)|^2$ to $r^2 /
(\pi^{3/2} \sigma^3) \exp(-r^2 / \sigma^2)$ with fitting parameter
$\sigma$, which is an appropriate fitting in 3D.
Figure~\ref{f:diff} shows that the numerically exact wave functions
significantly deviate from the Gaussian functions even in the stable
droplet.
We note that the Gaussian function in Fig~\ref{f:diff} (a) is almost the
same as that in Fig~\ref{f:diff} (b).
This indicates that the outer region, which has large weight in the
fitting due to the factor $r^2$, is almost stationary, and the mass flow
occurs mainly around the center of the droplet.
Thus, more appropriate functions are needed to correctly describe the
phenomena.

\section{Conclusions}
\label{s:conc}

We have studied a Bose-Einstein condensate with an oscillating interaction
with dissipation, and shown that a BEC droplet is stabilized in 3D free
space with realistic dissipation, $\gamma = 0.03$.
The frequency of the oscillating interaction must be much faster than the
characteristic frequency of the system, but it can be a moderate frequency
according to the scaling property discussed in Sec.~\ref{s:GP}.
For instance, the situation in Fig.~\ref{f:stable} can be realized in a
condensate of $10^4$ ${}^{85}{\rm Rb}$ atoms by the s-wave scattering
length $a(t) = -0.6 + 1.35 \sin \Omega t$ [nm] with $\Omega = 100 \times 2
\pi$ Hz, where a trapping potential with frequency 3.3 Hz is removed in 25
ms, and the final size of the droplet becomes a few micrometers.
Such an oscillation of interaction can be easily realized in experiments
using the Feshbach resonance~\cite{Claussen}.

Under gravity, a condensate falls after the trapping potential is turned
off.
In the above example, the gravitational sag in the initial trapping
potential is $\simeq 2$ cm, and the condensate falls about 1 mm until the
trapping potential vanishes.
The effect of gravity can be canceled out using the magnetic
levitation~\cite{Herbig}, which enables us to observe the longtime
behavior of the droplet.

A ``gaseous BEC droplet'' found in this paper is different from the usual
condensate in that it coheres by itself without the help of the trapping
potential.
This self-trapped matter wave might exhibit interesting dynamics, such as
in collective mode, collapsing dynamics, and vortex nucleation, which will
be discussed elsewhere.

{\it Note added} --- While preparing the present paper, a similar
paper~\cite{Adhikari} appeared which also reports stabilization of a BEC
in 3D free space by an oscillating interaction.
The strength of interaction studied in Ref.~\cite{Adhikari} is much larger
than that of ours.
In our parameter regime, a droplet can be stabilized adiabatically by
simple ramp functions with realistic dissipation.
On the other hand, in Ref~\cite{Adhikari} a droplet is stabilized without
dissipation by following a complicated ramp scheme to reach a final
state.

\begin{acknowledgments}
This work was supported by the Special Coordination Funds for Promoting
Science and Technology, a 21st Century COE program at Tokyo Tech
``Nanometer-Scale Quantum Physics'', and a Grant-in-Aid for Scientific
Research (Grant No. 15340129) from the Ministry of Education, Science,
Sports, and Culture of Japan.
\end{acknowledgments}


\begin{thebibliography}{}
\bibitem{note}
We call a self-trapped BEC in 3D a ``droplet'' because, strictly
speaking, it is not a ``soliton''.
\end{thebibliography}
\end{document}